# Observation of topological surface phonons in diamond with nonlinear optics

Qi Wang[1,2,#], Xinyi Liu[3,#], Xinyue Sheng[3,#], Zhi-Kang Lin[1,4,5,#], Xiaowei Lu[1,2,†], Xiaosheng Yang[6], Peining Li[6], Yizhou Liu[7], Chang-Hui Li[8,9], Wei-Tao Liu[3,†], and Jian-Hua Jiang[1,2,10,†]

[1]*State Key Laboratory of Bioinspired Interfacial Materials Science, Suzhou Institute for Advanced Research, University of Science and Technology of China, Suzhou 215123, China*

[2]*School of Biomedical Engineering, Suzhou Institute for Advanced Research, University of Science and Technology of China, Suzhou 215123, China*

[3]*Physics Department, State Key Laboratory of Surface Physics, Key Laboratory of Micro and Nano Photonic Structures [Ministry of Education (MOE)], Fudan University, Shanghai 200433, China*

[4]*Institute for Quantum Science, School of Physical Science and Technology, Soochow University, Suzhou 215006, China*

[5]*Department of Physics, The University of Hong Kong, Pokfulam Road, 999077, Hong Kong*

[6]*Wuhan National Laboratory for Optoelectronics & School of Optical and Electronic Information, Huazhong University of Science and Technology, Wuhan, 430074, China*

[7]*Center for Phononics and Thermal Energy Science, China-EU Joint Lab on Nanophononics, Shanghai Key Laboratory of Special Artificial Microstructure Materials and Technology, School of Physics Science and Engineering, Tongji University, Shanghai, 200092, China*

[8]*Lishui Spring Technology Co., Ltd., Longquan 323700, Zhejiang Province, China*

[9]*Longquan Industrial Innovation Research Institute of Zhejiang University, Longquan 323700, Zhejiang Province, China*

[10]*School of Physical Sciences, University of Science and Technology of China, Hefei 230026, China*

[#]These authors contributed equally to this work.

[†]Correspondence should be sent to: jhjiang3@ustc.edu.cn (Jian-Hua Jiang), wtliu@fudan.edu.cn (Wei-Tao Liu), or xwlu@ustc.edu.cn (Xiao-Wei Lu).

## Abstract

**Topological quantum states in electronic systems have profoundly transformed the understanding of phases of matter. Recent theories predict novel vibrational topological quantum states, i.e., topological phonons in various solids. However, this paradigm is yet to be established due to the lack of convincing experimental verification of topological surface phonons—a hallmark signature of topological phonon states. Here, we report the discovery of topological surface phonons in diamond using sum-frequency spectroscopy—a nonlinear optical spectroscopy capable of probing surface phonons with high sensitivity. Diamond, known for exceptional hardness and thermal transport, is unveiled as a phonon topological semimetal hosting topological nodal-lines and nexus triple points in the bulk. With consistent theory and experiments, we uncover the resultant topological surface phonons on diamond (111) and (100) surfaces. Moreover, by chemically modifying these surfaces, we reveal the disorder effect on topological surface phonons. These findings pave the way for bridging two fundamental domains: quantum topology and lattice dynamics, besides having important implications on diamond-based devices and functional interfaces.**

## Introduction

Phonons, the quanta of lattice vibration, are fundamental quasiparticles in solids, playing a pivotal role in phenomena such as electrical and thermal transport, quantum dissipation, carrier dynamics, optical properties, as well as superconductivity. Their interactions with electrons and photons underpin diverse effects, including Raman scattering, phonon polaritons, charge density waves, and Cooper pair formation, to name but only a few.

Over the past years, topological quantum states has been generalized to phonons (vibrational states) in solids (Figs. 1a-b) [1-5], giving rise to rich topological phonon states including Weyl phonons [6-9], Dirac phonons [10-13], nexus triple phonons [14], phonon nodal-lines [14-16], and non-Abelian topological phonon states [17,18]. In fact, topological phonons are found to exist ubiquitously in various solid-state materials [4,5,14]. They are also connected to thermal Hall effect [19,20], chiral phonons [21-24], and other intriguing phenomena. However, the hallmark of topological phonon states—topological surface phonons (TSPs) arising from bulk-surface correspondence have not yet been confirmed in experiments. Although in mechanical metamaterials topological edge and surface phonons have been explored [25-27], these phonons, however, have much lower frequency and are in macro- or mesoscopic scales where materials are regarded as continuous media instead of discrete lattices of atoms, and therefore have very different physics. Moreover, there is no quantization involved in these low-frequency phonons when they interact with electrons and other particles. Thus, they are instead regarded as classical waves.

Diamond, combining exceptional hardness, chemical inertness, a wide electron bandgap, nitrogen-vacancy centers, and unparalleled thermal conductivity together, is a key material in electronics, electrochemistry, quantum information science, and thermal management [28,29]. Studying the vibrational topological states (topological phonons) in diamond offers valuable insights into the fundamental physics dictating phonon-related properties and opens new ways to optimizing applications based on diamond. For instance, it was observed that hydrogen-terminated diamond surfaces offer high-quality *p*-channels on top of the wide-gap insulating bulk [30], which is promising for, e.g., field-effect transistors operating at high voltages and frequencies. Through electron-phonon couplings, TSPs can play a notable role in understanding charge transport in those surface *p*-channels.

However, probing TSPs in solid materials remains challenging due to intrinsic and technical limitations. First, phonons' energy scale is too small (~ tens of meV)—comparable with or even lower than the energy resolution of electron spectroscopy [5]. Second, the response of surface phonons is often significantly weaker than that of bulk phonons in spectroscopy, while they often mix together. High-resolution electron energy loss spectroscopy is valuable for measuring phonon dispersions [13], despite facing practical challenges such as limited energy resolution, low sensitivity, and almost inapplicable to insulators. Furthermore, in many cases, TSPs can be infrared or Raman inactive, restricting studies based on conventional optical spectroscopy. To date, detection of topological phonons is still focused on bulk phonons, while probing TSPs remains a prominent experimental challenge (Fig. 1a).

Sum-frequency spectroscopy (SFS) is a nonlinear optical spectroscopy featured with exceptional sensitivity especially in detecting surface phonons, besides having much higher energy resolution [31]. When applied to centrosymmetric materials, SFS directly detects surface phonons without interacting with the bulk phonons (Fig. 1b). Furthermore, through polarization anisotropy of SFS, symmetry properties of surface phonons can be analyzed. Therefore, SFS is an excellent tool for investigating TSPs in centrosymmetric, insulating materials such as diamond where we find TSPs due to, e.g., phonon nodal lines (Fig. 1a) and nexus triple points—unique topological phonon states in diamond and similar crystals [14].

# Results

## Theory

Diamond (space group No. 227, Fig. 1c) hosts rich topological degeneracies in phonon bands, including topological nodal-lines and nexus triple points, as revealed by *ab-initio* calculations (Fig. 1d). In this work, we focus on the nodal-lines due to accidental degeneracy (red arrow, see SI for details). Meanwhile, nexus triple points emerge when the LA band crosses the degenerate TO bands on the Γ-X line. Phonon bands calculation here agrees well with the inelastic neutron and X-ray scattering data in Refs. [32,33]. This agreement confirms the bulk topological phonons, i.e., the phonon nodal-lines and nexus triple points, in diamond. Intriguingly, the topology of nodal-lines and nexus triple points are manifested on distinct surfaces, (111) and (001) surfaces, respectively, due to the bulk-surface correspondence (see SI for more details).

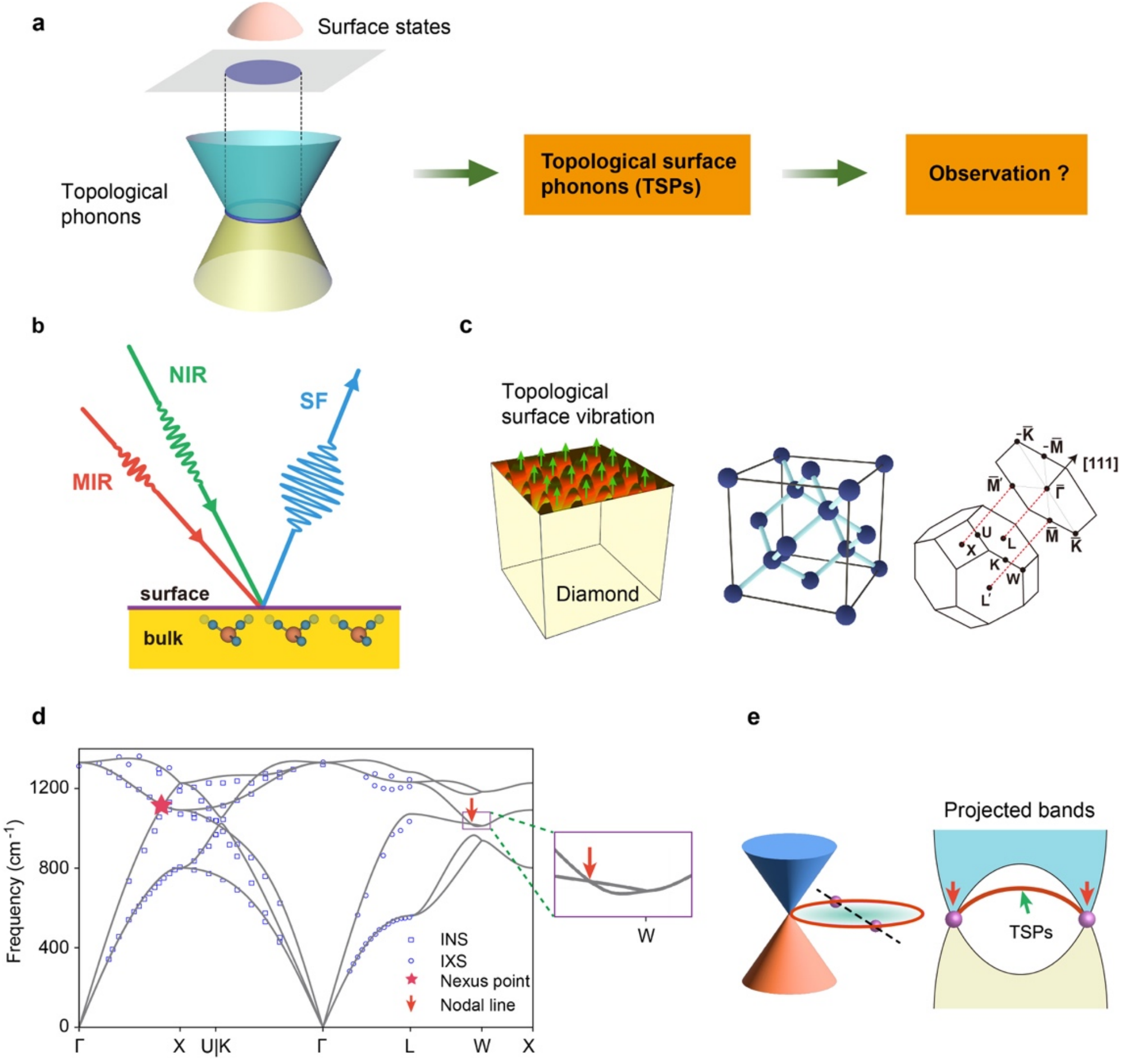

**Figure 1 | Phononic topological states. a,** Phonon topology and its challenge. Left: Illustration of topological phonons forming a nodal line and the resultant topological surface states. Right: Emergence of topological surface phonons (TSPs) and the experimental challenge of observing

them. **b,** Surface phonons can be detected by SFS which is surface-specific and highly sensitive. **c**, Schematic of topological surface vibration as the manifestation of TSPs in diamond which hosts topological phonons (Left). Right: Diamond's atomic structure and its Brillouin zones for the bulk and the (111) surface. **d**, Phonon band structure of diamond. Curves: calculation; dots: experiments from inelastic neutron scattering (INS) and inelastic X-ray scattering (IXS). A nexus triple point and a nodal line are marked, respectively, by a star and an arrow. Inset: zoom-in on the phonon dispersion around the nodal line. **e**, Schematic of a topological nodal-line (left) and the projected phonon dispersion onto a surface along the dashed line direction (right).

The topological properties of these nodal-lines and nexus triple points can be revealed by the symmetry eigenvalues of phonon bands or directly calculating the Berry phases of these bands (see SI). Notably, the degeneracy, linking, and topology of the phonon bands are largely determined by the crystalline symmetry of diamond, manifesting an elegant case of symmetry-protected topological quantum states.

A topological nodal-line is characterized by Berry phase $\pi$ on a loop encircling it. Each point on the nodal-line can be extended to a Dirac cone in energy-momentum space (Fig. 1e) [34]. The surface local density-of-states along, e.g., the dashed line will give TSPs as a Dirac string connecting the projected nodal-lines (the red arrows). As shown in Fig. 2a, TSPs emerge in the spectral gap of bulk phonons and connect to the phonon nodal-lines, manifesting the bulk-surface correspondence as schematized in Fig. 1d (see SI for more evidences).

To reveal the nature of these TSPs, we obtain their vibrational patterns from *ab-initio* calculations by constructing a finite-thickness supercell (see SI for details). The results for (111) surfaces give three types of surface phonons with distinct symmetry and spatial distributions (Figs. 2b-2d): (1) Trivial surface phonons associated with vibrations of dangling bonds that are strongly confined in the outermost atomic layer and characterized by in-plane vibrations and frequencies even higher than bulk phonons (Here, in- and out-of-plane are defined with respect to the surface plane and its parallel planes); (2) Bulk-derived surface phonons that are shallow surface modes with much extended vibration patterns, carrying both in- and out-of-plane vibrations; and (3) TSPs with vibrations extending over several atomic layers beneath the surface, featured by out-of-plane vibrations. Besides, their frequency at the Brillouin zone center, 1160 $cm^{-1}$, falls right in the spectral gap of bulk phonons. These properties, distinct from other surface phonon modes, are consistent with previous studies on TSPs [14-16]. Further, calculations indicate that trivial surface phonons are sensitive to disorders, whereas the TSPs are robust against disorders (see SI).

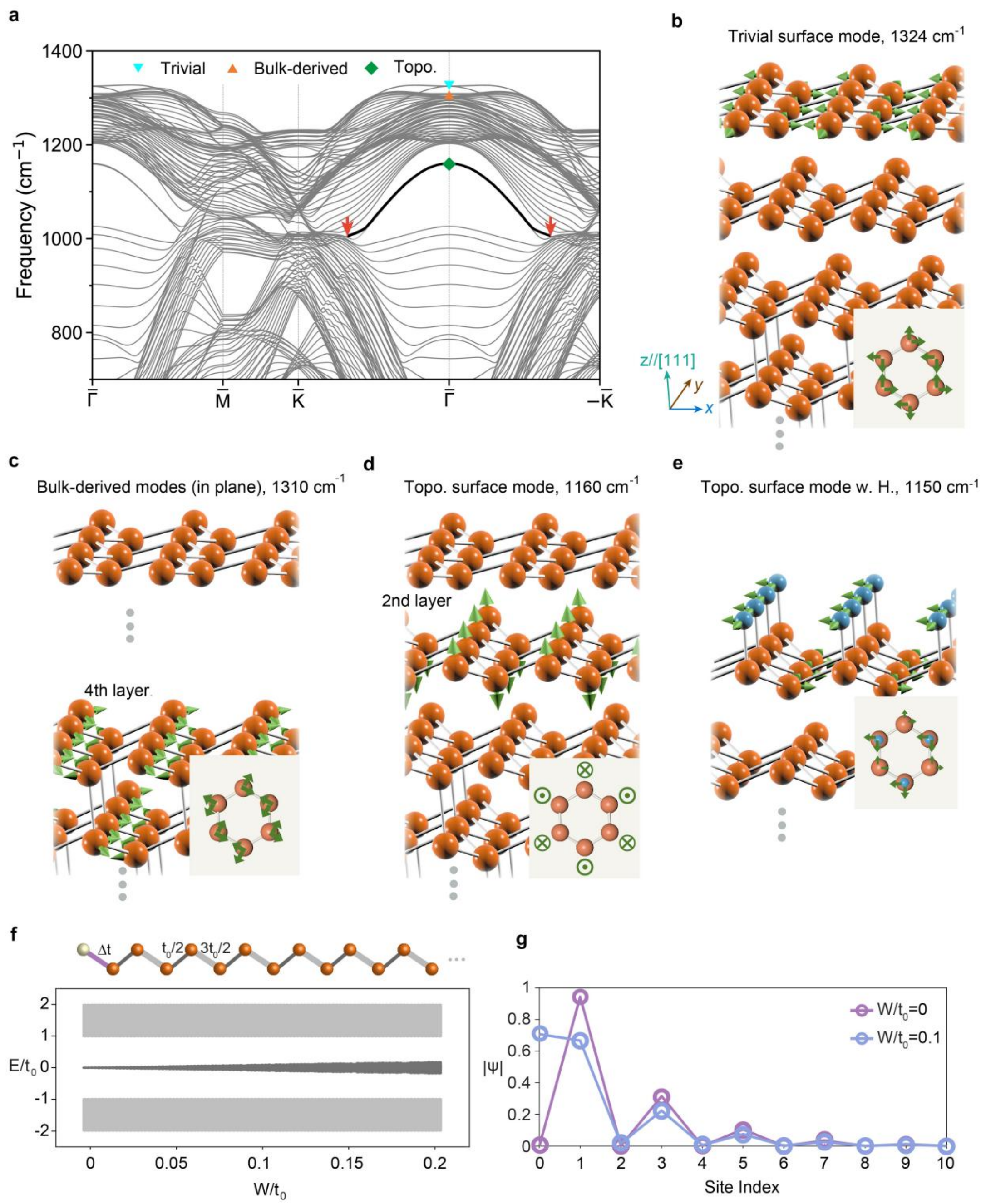

**Figure 2 | Topological phonons on diamond (111) surfaces.** **a**, Bulk and surface phonon dispersions for diamond (111) surfaces along the high-symmetry lines of the surface Brillouin zone. The trivial, bulk-derived, and topological surface phonons at surface Brillouin zone center is marked by various symbols. TSPs' dispersion is emphasized by the thick black curve. Orange arrows mark the projection of the nodal lines in the surface Brillouin zone. **b-d,** Illustration of the vibration geometry for the trivial (**b**), bulk-derived (**c**), and topological (**d**) surface phonons where arrows stand for the atomic displacements. The coordinate system is depicted in **b**. Insets present the atomic vibrations from a top-down view. **e**, Schematic of the vibration pattern for topological surface phonons on a fully-hydrogenated diamond (111) surface. Orange and blue spheres denote carbon and hydrogen atoms, separately. **f-g**, Su-Schrieffer-Heeger model with

surface perturbations where the intra- and inter-unit-cell couplings are $t/2$ and $3t/2$, respectively. **f**, Energy spectrum of the model when surface perturbation $\Delta t$ (schematized on top) is randomly distributed in the region [-W, W] (uniform random distribution). **g**, Wavefunctions of the edge states without and with the surface perturbation.

As diamond is a centrosymmetric crystal, sum-frequency (SF) generation arises exclusively from surface phonons when higher-order corrections are neglected. This directly eliminates bulk phonons' contribution and leads to convincing evidence for TSPs. Furthermore, the SF signal is proportional to the dipole derivative which can also be obtained from the *ab-initio* calculations. Calculations (see SI) show that in a broad frequency window, only two SF resonances emerge that are due to the TSPs and bulk-derived surface phonons, which is confirmed later in our experiments. The bulk-derived surface phonons are in fact surface phonons loosely confined to the surfaces that gradually evolve into bulk-like phonon dispersions in thick films as confirmed in subsequent experiments (see SI for more details).

*Ab-initio* calculations can also reveal the influence of surface chemical bonds on surface phonons. Interestingly, even if the surfaces are fully hydrogenated (i.e., covered completely with surface C-H bonds), the TSP undergoes only a small frequency redshift from 1160 cm$^{-1}$ to 1150 cm$^{-1}$ (Fig. 2e). However, these surface C-H bonds can induce prominent changes in vibration geometry (Fig. 2**e**). This indicates that surface chemical bonding has a stronger effect on the vibrational geometry than on the frequency of TSPs. This phenomenon is consistent with existing theories [38,39]: With surface disorder or perturbations, the wave patterns of topological edge states are deformable, but their spectral properties are more stable. This property can be illustrated using a perturbed Su-Schrieffer-Heeger model as an example (see Figs. 2**f**-**g**). In the model, an additional site is attached to the left edge with a random coupling $\Delta t$ to emulate the effect of the random surface chemical bonds. Results show that such perturbation can considerably change the wavefunction of the edge state, while affecting its energy marginally [38,39].

## Experiments: Measuring topological surface phonons on (111) surfaces

Experimental setup for SFS is shown in Fig. 3a (left): When a near-infrared (NIR) laser (frequency $\omega_1$) and a mid-infrared (MIR) laser (frequency $\omega_2$) overlap at a spot on a diamond surface, SF radiation is emitted at the frequency $\omega_3 = \omega_1 + \omega_2$. The two input lasers are generated by the same laser system (from Spectra-Physics Inc.) using different ways. The SF signal is proportional to the square of the absolute value of the second-order nonlinear optical susceptibility tensor, $\chi^{(2)}$, providing a direct measure of the surface phonons. The quantum process is illustrated in Fig. 3a (right): the MIR photon excites a phonon state ($|q\rangle$) from the ground state ($|g\rangle$) which is upconverted by the NIR photon to a virtual state $|v\rangle$ and then converted back to the ground state by emitting an SF photon. Upon hitting a surface phonon state $|q\rangle$, the nonlinear optical tensor $\chi^{(2)}$ is resonantly enhanced with the resonant part given by $\chi_R^{(2)} \propto \frac{A_q}{\omega_2 - \omega_q + i\gamma_q}$. Here, $A_q \propto d_q R_q$ with $A_q$, $\omega_q$, $\gamma_q$, $d_q$, and $R_q$ denoting the resonance amplitude, frequency, damping coefficient, dipole derivative, and Raman polarizability of the surface phonon, respectively [31] (see SI for details).

Starting with a commercial (111) diamond with an initial root-mean-square surface roughness of 3.1 nm, our sequential surface treatments—acid etching, annealing, and hydrogen plasma exposure (see Methods)—substantially reduce the surface roughness to 0.12 nm. The crystal and surface structures are characterized by X-ray diffraction, atomic force microscope (see Fig. 3b), and low-energy electron diffraction (see SI). All these characterizations show that our samples are high-quality diamond crystals with excellent (111) surfaces. Diamond surfaces often contain C-H and C-O chemical bonds at ambient condition [40-42]. With hydrogen plasma exposure, our samples are initiated with negligible C-O bonds but rich C-H bonds (see X-ray photoelectron spectroscopy in Fig. 3c).

In SFS, polarizations of the SF, NIR, and MIR beams can be varied (i.e., *s*- or *p*-polarized), accessing different configurations that are sensitive to the spatial alignment of surface-phonon-induced dipoles. With PPP configuration (all beams *p*-polarized), which couples to both in-plane and out-of-plane dipoles, our measurements for (111) surfaces reveal two well-separated peaks at 1190 $cm^{-1}$ and 1334 $cm^{-1}$, respectively (Fig. 3d). These peaks are also observed for other polarization configurations (e.g., SSP). Remarkably, these peak frequencies align well with the calculated topological and bulk-derived surface phonons, with minor shifts ~30 $cm^{-1}$ (i.e., ~3.7 meV) attributable to the intrinsic error of *ab-initio* calculation for phonons in finite-size supercells. Importantly, both peaks are spectrally separated from the vibrational modes of surface adsorbates or molecular contaminants on diamond surfaces [42]. Together with the strong SF signals, this ensures that the observed SFS resonances originate from intrinsic surface phonons in diamond. The 1334 $cm^{-1}$ resonance, very close to the measured Raman resonance at 1333.6 $cm^{-1}$ of bulk optical phonons (Fig. 3e), indicates that it should be associated to the bulk-derived surface phonons. In contrast, the 1190 $cm^{-1}$ resonance is within bulk-phonons' projected spectral gap and can thus only be assigned as the TSPs.

We emphasize that the Raman scattering cannot probe the TSPs (see Fig. 3e: no resonance from 1000 to 1300 $cm^{-1}$), whereas SFS can achieve such a goal—manifesting the special sensitivity for surface phonons of SFS. Further, we find in experiments that the PPP polarization gives stronger resonance than the other polarizations for the TSPs in SFS. This observation indicates that the dipole moment of the TSPs is predominantly along the out-of-plane direction, which is consistent with the ab initio calculation results in Fig. 2. With such a dipole moment, TSPs cannot be probed by conventional infrared optical spectroscopy, as confirmed in our experiments (see SI).

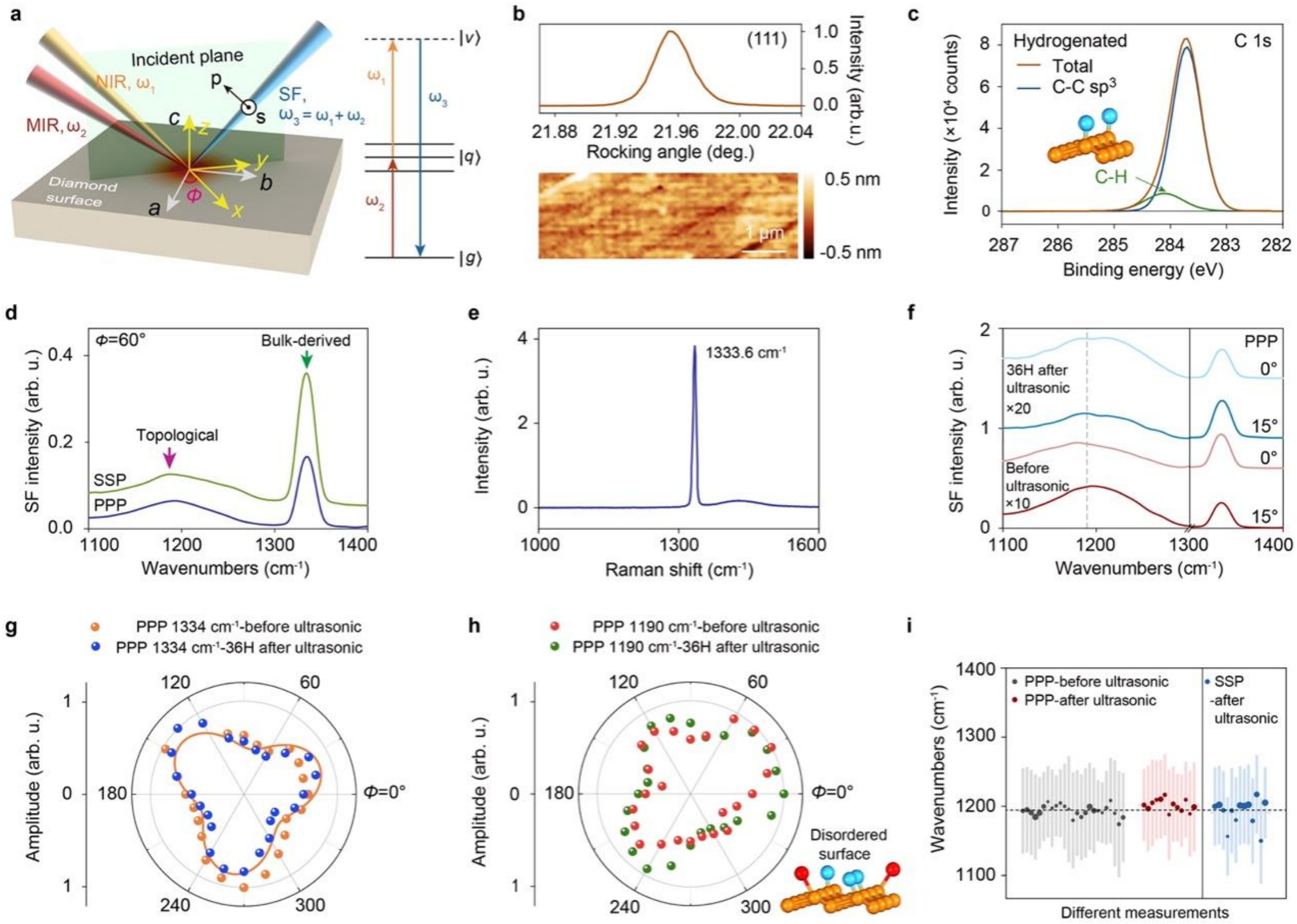

**Figure 3 | Probing topological surface phonons on diamond (111) facets. a,** Schematic of SFS of a diamond surface (left). Two laser beams in near-infrared (NIR) and mid-infrared (MIR) frequency ranges, respectively, overlap at a surface spot to induce SF generation. Right panel: quantum process for SF generation. **b**, X-ray diffraction (upper) and atomic force microscope surface topography (lower) of a (111) diamond wafer. **c**, X-ray photoelectron spectroscopy for a diamond (111) surface right after hydrogen plasma exposure, showing peaks arising from different surface chemical bonds. **d**, SFS in PPP and SSP configurations for a (111) surface (spectra vertically displaced for clarity). **e**, Raman spectroscopy for a (111) diamond wafer. **f**, SFS before (red) and 36 hours after (blue) ultrasonic treatment (spectra vertically displaced for clarity) for a (111) surface. Note that data below 1300 $cm^{-1}$ have been rescaled by 10 and 20 times for the red and blue curves, respectively, to be more noticeable. **g-h**, SFS anisotropy at 1334 $cm^{-1}$ (**g**) and 1190 $cm^{-1}$ (**h**) before and 36 hours after the ultrasonic treatment. Dots denote experimental data. Curve gives a three-fold anisotropy fitting of the data. **i**, Graphical summary of peak frequency and broadening of the SFS resonance associated with topological surface phonons at different measurements (at different time or laser-spot position). Dots give the peak frequency, while the dot size represents the peak's height. Shades denote the peaks' full width

at half maximum. Colors represent different measurements. Dashed line stands for the peak frequency 1195 $cm^{-1}$ averaged over all data.

With high surface-sensitivity of SFS, we can further study the effects of disorder on the surface phonons. Here, we exploit the surface chemical bonds as a resource of disorder. In particular, our samples are deliberately hydrogenated and are rich in C-H bonds (Fig. 3c) that are randomly distributed on the surfaces. These and other random surface bonds, together with surface roughness, create strong, tunable surface disorders that can be utilized to investigate how surface disorder affect various surface phonon states that are predicted in Fig. 2. An inevitable consequence here is, however, that the SFS resonance associated with the TSPs is much more broadened than that of the bulk-derived surface phonons. We find that this is a common feature of TSPs, essentially because in reality, crystal surfaces are significantly more disordered than the bulk.

By further modifying the surfaces with surface treatments, we can tune the surface disorder and study its effect on vibrational and spectral properties of surface phonons. As an advantage of SFS, these properties are contained in SFS: The spectrum is given by SFS resonances, while the vibrational properties are reflected by the SFS anisotropy which is tied to the symmetry of the phonon-induced dipole. In our experiments, SFS anisotropy, i.e., azimuthal angle $\phi$-dependence of SF signal (Fig. 3a), is measured by fixing the beams but rotating the samples.

To tune the surface chemical bonds, we use various surface treatments, including UV irradiation, $D_2O$ and $H_2O$ cleaning, as well as ultrasonic treatment (see Methods), which can remove part of the C-H bonds and adsorbates on the surfaces. Besides, under ambient conditions, C-H bonds on diamond surface will be gradually oxidized [28,43] (see XPS evidence in SI), which also changes the surface disorder.

We find that both the resonant frequency and the anisotropy of the resonance at 1334 $cm^{-1}$ is quite stable under surface treatments (Figs. 3f-g). These results indicate that bulk-derived surface phonons are quite insensitive to surface disorder, which is consistent with that their vibrations considerably extend away from the surface (Fig. 2b and SI). This bulk-like nature is also reflected by the three-fold anisotropy of SFS which agrees with the three-fold rotation symmetry of (111) diamond. In comparison, TSPs can be notably affected by surface disorder and surface treatments (Fig. 3f, see SI for more data). Here, we discussed only the ultrasonic treatment and the gradual oxidization afterwards, while we find that other treatments have similar effects (SI).

Unlike the resonance at 1334 cm⁻¹, SFS anisotropy at 1190 cm⁻¹ is rather irregular (Fig. 3h) which is incompatible with the symmetry of (111) diamond and can only be assigned to strong surface disorder. Especially, the random distribution of C-H bonds and their interaction with surface defects and other bonds can mess up the surface-phonon-induced dipole moments [42] and lead to irregular SFS anisotropy.

We further find that due to gradual surface oxidization, the measured SFS slowly changes with time and laser-spot position. Remarkably, when putting all measurements for different time, laser-spot positions, and surface treatments together, we find that the data give quite consistent results (Fig. 3i): A stable resonance peak around 1190 cm⁻¹ (with overall averaged peak frequency 1195 cm⁻¹) confirming the predicted TSPs emerging on (111) surfaces due to the bulk topological phonon nodal-lines. We remark that all data in Fig. 3i, including the broadening, fall precisely into the projected bulk spectral gap from 1067 cm⁻¹ to 1227 cm⁻¹, confirming that these peaks must result from the topological surface phonons.

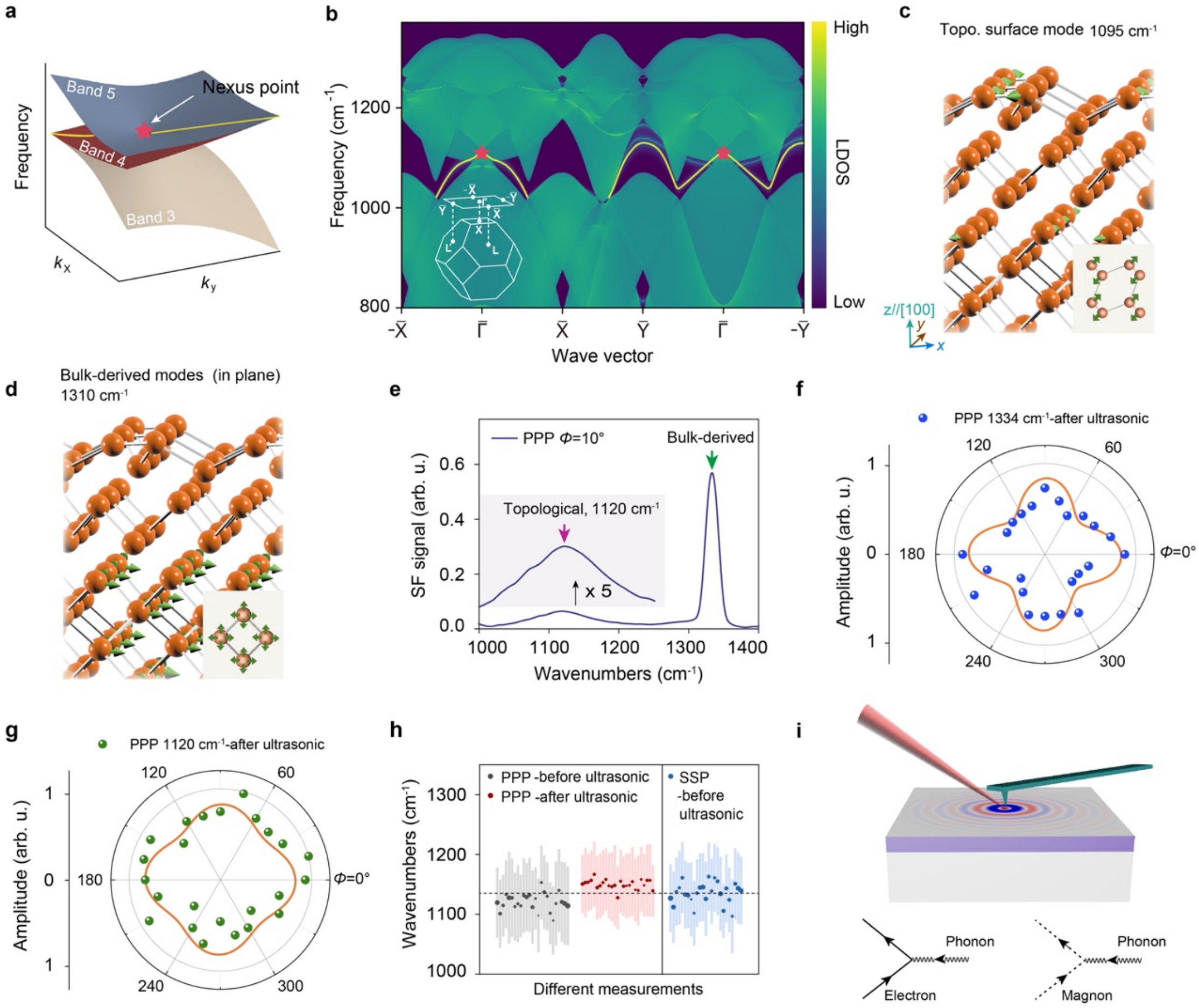

**Figure 4 | Topological surface phonons on (100) facets. a**, Illustration of a nexus triple point in phonon dispersion. **b**, Phonon LDOS on a diamond (100) surface. Stars and yellow curves represent the nexus triple points and associated topological surface states, respectively. Inset

gives the bulk and surface Brillouin zones. **c-d**, Illustration of the vibrational geometry for topological (**c**) and bulk-derived (**d**) surface phonons. Arrows stand for atomic displacements. The coordinate system is depicted in **c**. Insets give the atomic vibrations from a top-down view. **e**, SFS in PPP configuration for a diamond (100) surface. **f**, Anisotropy of SF signals at 1334 $cm^{-1}$ for the bulk-derived surface phonons. **g**, Anisotropy of SF signals at 1120 $cm^{-1}$ for the topological surface phonons. **h**, Graphical summary of the peak frequency and broadening of the SFS resonance associated with topological surface phonons from different measurements (i.e., measured at different times or laser-spot positions). Dots give the peak frequency, while the dot size stands for the peak's height. Shades give full width at half maximum of the peaks. Colors stand for different measurements. Dashed line denotes the averaged peak frequency 1135 $cm^{-1}$. **i**, Schematic of potential future research. Top: Tip-enhanced light-matter interaction enabled by topological surface phonons. Down: Interaction and scattering between topological surface phonons and electrons or magnons at functional interfaces.

## Theory and experiments for topological surface phonons on (100) surfaces

We now use both theoretical calculation and SFS measurements to investigate the topological phonon states on (100) surfaces due to nexus triple points. A nexus triple point is featured with non-Abelian multi-band topology [41-43]. In diamond, such triple degeneracy has higher codimensions and can only be a point that connects two nodal sections along high-symmetry lines, e.g., the Γ-X line (see Figs. 1d and 4a; In Fig. 4a, the Γ-X line is repsented by the yellow line). It is an accidental degeneracy between the doubly degenerate TO bands and the LA band (Fig. 1d). A nexus triple point can be regarded as the coalesce of two Weyl points with opposite topological charges [14,37]. However, these Weyl points cannot annihilate each other as they come from different band sets. Here, one Weyl point is formed by the third and fourth bands, while the other is formed by the fourth and fifth bands (Fig. 1e). The nexus triple points are stable and characterized with nontrivial multi-gap non-Abelian topology [14,37] that leads to the Fermi-arc-like surface states as shown in Fig. 4b (see Methods for LDOS calculations and SI for more details).

Finite-size supercell calculation gives the TSPs near the nexus triple point with a frequency of 1095 $cm^{-1}$ and strong localization near the surface (Fig. 4**c**), contrasting sharply with the bulk-derived surface phonons around 1310 $cm^{-1}$ which have vibration patterns extending notably away from the surface (Fig. 4**d**). We calculate the dipole derivative of these surface phonons and find that the dipole moments of TSPs are mainly in the (100) plane (see SI). In comparison, the bulk-derived surface phonons have comparable in-plane and out-of-plane dipole moments. We further confirm that with surface disorder, both topological and bulk-derived surface phonons survive (see calculation results in SI). Specifically, with surface C-H bonds, TSPs are only slightly red-shifted (see SI).

We measure the SFS of a (100) diamond single crystal wafer of which the root-mean square surface roughness is 0.4 nm (see SI). The X-ray diffraction curve agrees well with that of a standard (100) diamond wafer (see SI), indicating that the wafer and the surfaces are of high quality. For both PPP (Fig. 4**e**) and SSP (see SI) configurations, SFS reveals two resonant peaks: one around 1120 $cm^{-1}$ attributed to TSPs, the other at 1334 $cm^{-1}$ corresponding to bulk-derived surface phonons. Here, the measured frequencies of TSPs are slightly blue-shifted by ~25 $cm^{-1}$ (i.e., ~3.1 meV) compared with *ab-initio* calculations (Figs. 4**c**-**d**), which can be assigned to the intrinsic variation in finite-size calculations.

In SFS resonances, the peak associated to bulk-derived surface phonons at 1334 $cm^{-1}$ has a four-fold anisotropy (see Fig. 4**f**) as dictated by the crystal surface symmetry inherited by these surface phonons. The SFS resonance for TSPs at 1120 $cm^{-1}$ also has a four-fold-like anisotropy (Fig. 4**g**) which may be due to (100) surface's reconstruction. In fact, it was known that diamond (100) surfaces exhibit a stable (2×1) reconstruction with dimerization along either the [010] or [001] direction [42,44]. For real-world (100) surfaces, these two types of surface reconstruction are randomly distributed with nearly equal probability, which makes the SFS anisotropy four-fold-like. This effect could be further enhanced by the surface C-H bonds which tilt the dipole moment of TSPs and make SFS anisotropy more four-fold-like (see SI).

Like that for (111) surfaces, although the SFS of TSPs depend on the measurement time, laser-spot position, and surface treatment, by putting the results together for different measurements, all data show a consistent feature (Fig. 4**h**): A resonance peak around 1120 $cm^{-1}$ arises from the TSPs (with overall averaged peak frequency 1135 $cm^{-1}$). A noticeable trend here is surface oxidization leads to frequency blue-shift of these surface phonons which is attributed to the influence of the surface C-O bonds (see SI). Overall, these observations confirm the predicted TSPs on diamond (100) surfaces due to the bulk topological nexus triple points.

## Conclusion and outlook

We discover topological phonons in diamond by consistent theory, calculations, and experiments. While the bulk topological phonons are revealed by comparing our calculations with previous neutron and X-ray scatterings data, the TSPs are observed for the first time by surface-sensitive SFS. These progresses unveil a new class of surface vibration quantum states—TSPs and set a cornerstone for bridging the two fundamental realms: quantum topology and lattice dynamics.

Moreover, our discovery has important implications for a range of cutting-edge applications. Going beyond diamond, as schematized in Fig. 4i, TSPs can interact with photons (via tip-enhanced light-matter interaction) as well as with electrons and magnons at interfaces. Since topological phonons are found to exist ubiquitously in solids [4, 14], and because of their large LDOS at interfaces, TSPs can have profound influence on material interfaces and functional devices [45]. For instance, interface scattering plays a major role in reducing the quality of magnetic junctions that are key for magnetic memory. Scattering between TSPs and magnons (or spin-polarized electrons) can play a notable role in spin relaxation across interfaces [46], especially the TSPs associated with the acoustic phonon nodal lines which have much lower frequencies and scatter more efficiently with electrons at ambient temperatures [45]. Moreover, TSPs could be important for understanding interfacial thermal resistance and heat transport at nanoscales, as demonstrated in Ref. [45].

## References

1 Po, H. C., Bahri, Y. & Vishwanath, A. Phonon analog of topological nodal semimetals. *Phys. Rev. B* **93**, 205158 (2016).

2 Liu, Y., Xu, Y. & Duan, W. Berry phase and topological effects of phonons. *Natl. Sci. Rev.* **5**, 314-316 (2018).

3 Liu, Y., Lian, C. S., Li, Y., Xu, Y. & Duan, W. Pseudospins and topological effects of phonons in a Kekulé lattice. *Phys. Rev. Lett.* **119**, 255901 (2017).

4 Li, J. *et al.* Computation and data driven discovery of topological phononic materials. *Nat. Commun.* **12**, 1204 (2021).

5 Xu, Y. *et al.* Catalog of topological phonon materials. *Science* **384**, eadf8458 (2024).

6 Li, J. *et al.* Coexistent three-component and two-component Weyl phonons in TiS, ZrSe, and HfTe. *Phys. Rev. B* **97**, 054305 (2018).

7 Zhang, T. *et al.* Double-Weyl phonons in transition-metal monosilicides. *Phys. Rev. Lett.* **120**, 016401 (2018).

8 Miao, H. *et al.* Observation of double Weyl phonons in parity-breaking FeSi. *Phys. Rev. Lett.* **121**, 035302 (2018).

9 Peng, B., Hu, Y., Murakami, S., Zhang, T. & Monserrat, B. Topological phonons in oxide perovskites controlled by light. *Sci. Adv.* **6**, eabd1618 (2020).

10 Jin, Y., Wang, R. & Xu, H. Recipe for Dirac phonon states with a quantized valley Berry phase in two-dimensional hexagonal lattices. *Nano Lett.* **18**, 7755-7760 (2018).

11 Chen, Z. J. *et al.* Three-dimensional Dirac phonons with inversion symmetry. *Phys. Rev. Lett.* **126**, 185301 (2021).

12 Jin, Z. *et al.* Chern numbers of topological phonon band crossing determined with inelastic neutron scattering. *Phys. Rev. B* **106**, 224304 (2022).

13 Li, J. *et al.* Direct observation of topological phonons in graphene. *Phys. Rev. Lett.* **131**, 116602 (2023).

14 Liu, Y. *et al.* Ubiquitous topological states of phonons in solids: silicon as a model material. *Nano Lett.* **22**, 2120-2126 (2022).

15 Zhang, T. T. *et al.* Phononic helical nodal-lines with *PT* protection in $MoB_2$. *Phys. Rev. Lett.* **123**, 245302 (2019).

16 Wang, X. *et al.* Topological nodal-line phonons: recent advances in materials realization. *Appl. Phys. Rev.* **9**, 041304 (2022).

17 Peng, B., Bouhon, A., Monserrat, B. & Slager, R. J. Phonons as a platform for non-Abelian braiding and its manifestation in layered silicates. *Nat. Commun.* **13**, 423 (2022).

18 Peng, B., Bouhon, A., Slager, R. J. & Monserrat, B. Multigap topology and non-Abelian braiding of phonons from first principles. *Phys. Rev. B* **105**, 085115 (2022).

19 Zhang, L., Ren, J., Wang, J. S. & Li, B. Topological nature of the phonon Hall effect. *Phys. Rev. Lett.* **105**, 225901 (2010).

20 Saito, T., Misaki, K., Ishizuka, H. & Nagaosa, N. Berry phase of phonons and thermal Hall effect in nonmagnetic insulators. *Phys. Rev. Lett.* **123**, 255901 (2019).

21 Zhang, L. & Niu, Q. Chiral phonons at high-symmetry points in monolayer hexagonal lattices. *Phys. Rev. Lett.* **115**, 115502 (2015).

22 Zhu, H. *et al.* Observation of chiral phonons. *Science* **359**, 579-582 (2018).

23 Ueda, H. *et al.* Chiral phonons in quartz probed by X-rays. *Nature* **618**, 946-950 (2023).

24 Ishito, K. *et al.* Truly chiral phonons in α-HgS. *Nat. Phys.* **19**, 35-39 (2023).

25 Ma, G. C., Xiao, M. & Chan, C. T. Topological phases in acoustic and mechanical systems. *Nat. Rev. Phys.* **1**, 281-294 (2019).

26 Xue, H. R., Yang, Y. H. & Zhang, B. L. Topological acoustics. *Nat. Rev. Mater.* **7**, 974-990 (2022).

27 Zhang, X., Zangeneh-Nejad, F., Chen, Z.-G., Lu, M.-H. & Christensen, J. A second wave of topological phenomena in photonics and acoustics. *Nature* **618**, 687-697 (2023).

28 Fischer, A. E., Show, Y. & Swain, G. M. Electrochemical performance of diamond thin-film electrodes from different commercial sources. *Anal. Chem.* **76**, 2553-2560 (2004).

29 Zulkharnay, R. & May, P. W. Applications of diamond films: a review. *Funct. Diam.* **4**, 2410160 (2024).

30 Sasama, Y. *et al.* High-mobility p-channel wide-bandgap transistors based on hydrogen-terminated diamond/hexagonal boron nitride heterostructures. *Nat. Electron.* **5**, 37-44 (2022).

31 Shen, Y.-R. *Second harmonic and sum-frequency spectroscopy: basics and applications*. (World Scientific, 2023).

32 Warren, J. L., Yarnell, J. L., Dolling, G. & Cowley, R. A. Lattice Dynamics of Diamond. *Phys. Rev.* **158**, 805-808 (1967).

33 Burkel, E. *Inelastic Scattering of X-Rays with Very High Energy Resolution*. Vol. 125 61-64 (1991).

34 Burkov, A. A., Hook, M. D. & Balents, L. Topological nodal semimetals. *Phys. Rev. B* **84**, 235126 (2011).

35 Wu, Q., Soluyanov, A. A. & Bzdušek, T. Non-Abelian band topology in noninteracting metals. *Science* **365**, 1273-1277 (2019).

36 Bouhon, A. *et al.* Non-Abelian reciprocal braiding of Weyl points and its manifestation in ZrTe. *Nat. Phys.* **16**, 1137-1143 (2020).

37 Lenggenhager, P. M., Liu, X., Tsirkin, S. S., Neupert, T. & Bzdušek, T. From triple-point materials to multiband nodal links. *Phys. Rev. B* **103**, L121101 (2021).

38 Scollon, M. & Kennett, M. P. Persistence of chirality in the Su-Schrieffer-Heeger model in the presence of on-site disorder. *Phys. Rev. B* **101**, 144204 (2020).

39 Shi, X. *et al.* Disorder-induced topological phase transition in a one-dimensional mechanical system. *Phys. Rev. Res.* **3**, 033012 (2021).

40 Lee, S. T. & Apai, G. Surface phonons and CH vibrational modes of diamond (100) and (111) surfaces. *Phys. Rev. B* **48**, 2684-2693 (1993).

41 Chin, R. P., Huang, J. Y., Shen, Y. R., Chuang, T. J. & Seki, H. Interaction of atomic hydrogen with the diamond C(111) surface studied by infrared-visible sum-frequency-generation spectroscopy. *Phys. Rev. B* **52**, 5985-5995 (1995).

42 Enriquez, J. I. *et al.* Oxidative etching mechanism of the diamond (100) surface. *Carbon* **174**, 36-51 (2021).

43 Hutton, L. A. *et al.* Examination of the factors affecting the electrochemical performance of oxygen-terminated polycrystalline boron-doped diamond electrodes. *Anal. Chem.* **85**, 7230-7240 (2013).

44 Frauenheim, T. *et al.* Stability, reconstruction, and electronic properties of diamond (100) and (111) surfaces. *Phys. Rev. B* **48**, 18189-18202 (1993).

45 Su, Z. *et al.* Topological surface phonons modulate thermal transport in semiconductor thin films. *Mater. Today Phys.* **66**, 102165 (2026).

46 Watts, J. D. *et al.* Finite-size effect in phonon-induced Elliott-Yafet spin relaxation in Al. *Phys. Rev. Lett.* **128**, 207201 (2022).

47 Bradlyn, B. *et al.* Topological quantum chemistry. *Nature* **547**, 298-305 (2017).

48 Liu, X. Y. *et al.* Nonlinear optical phonon spectroscopy revealing polaronic signatures of the $LaAlO_3/SrTiO_3$ interface. *Sci. Adv.* **9**, eadg7037 (2023).

49 Cao, Y. *et al.* Evolution of anatase surface active sites probed by in situ sum-frequency phonon spectroscopy. *Sci. Adv.* **2**, e1601162 (2016).

# Methods

## Sample treatments and characterizations

(111)- and (100)-oriented diamond single crystals were provided by the Carbon Cube Semiconductor Co., Ltd., Xiamen, China. To remove contaminants and achieve well-ordered surfaces, the diamond wafers were first treated in a 3:1 mixture of sulfuric and nitric acid at 220 $^{o}$C for 30 minutes. The diamond crystals were then annealed at 1000 $^{o}$C for 10 minutes, followed by exposure to 300 W hydrogen plasma at 800 $^{o}$C for 1 hour, in a vacuum chamber. The diamond wafers' crystal quality was assessed using high-resolution X-ray diffraction (XRD, EMPYREAN, Malvern Panalytical). Surface roughness was measured by atomic force microscope (AFM, NTEGRA II, NT-MDT). Lattice vibrational modes of bulk diamond crystals (optical modes) were investigated by using a HORIBA LabRAM Soleil Raman spectrometer, with a 532 nm laser as the excitation source. X-ray photoelectron spectroscopy (XPS, ESCALAB Xi+, Thermo Fisher Scientific) was utilized to monitor the evolution of the carbon valence state, with the charge compensation function applied to eliminate the charging effect. Surface crystal structure in the reciprocal space was analyzed by low-energy electron diffraction (LEED, Scienta Omicron) at an incident electron energy of 150 eV. Diamond surface treatments after the first-round SFS measurements include a standard cleaning process involving sequential sonication in isopropanol, ethanol (analytical reagent; Shanghai Sinopharm Chemical Reagent Co., Ltd.), and deionized water (18.2 MΩ·cm), each for 15 minutes. The ultrasonic instrument was SB-120DT model (Ningbo Scientz Biotechnology Co., Ltd.). The cleaned sample was blown dry with pure nitrogen for SFS measurement. Prior to other SFS measurements, diamond wafers were treated with different surface procedures: the deposition and evaporation of a droplet of $H_2O$ (18.2 MΩ·cm) or $D_2O$ (J&K Scientific, 99.8%), or UV irradiation (365 nm, 10 minutes) to induce photochemical surface modifications.

## Theory and calculations

All *ab-initio* calculations are performed via Vienna Ab initio Simulation Package (VASP, version 6.4). The Green's function method is employed to obtain the surface local density of states for phonons in a semi-infinite geometry. Finite-sized calculations are employed to obtain the optical properties (including the dipole derivative and the Raman tensor). Structure relaxation has been taken into account for the finite size calculations. The spectrum and eigenvectors of phonon is obtained by diagonalizing the dynamic matrix. Born effective charges and dielectric tensors are calculated by the density functional perturbation theory. The dipole derivative and Raman tensor are calculated based on similar perturbation theories (see SI for details). The topological theory of the nodal-lines and the nexus triple points is established via the symmetry representation of phonon Bloch bands, the Wilson-loop approach, the Zak phases, and the topological band theory (including the homotopy theory and the band representation theory) [47]. Details of these theories are presented in the SI. Finally, we emphasize that at the heart of the theory is the dynamic matrix and the normal modes description of the lattice vibration waves. This wave dynamic picture is distinct from the tight-binding picture in conventional models for topological states [1].

**Laser experiment setup**

The experiment setup is as described in previous studies [48, 49]. The details are as follows: We used a regenerative amplifier (Spitfire Ace, Spectra-Physics Inc.) seeded by a Ti: Sapphire oscillator (MaiTai SP, Spectra-Physics Inc.) to produce ~7 W of 800 nm, 35 fs pulses at 2 kHz repetition rate. 40% of the near-infrared beam passed through a beam-splitter to pump an optical parameter amplifier (TOPAS-Prime, Spectra Physics Inc.) followed by a difference frequency generation stage. The rest of beam power was reflected through a Bragg filter (N013-14-A2, OptiGrate) to generate a narrowband beam of ~0.5 nm bandwidth. The broadband MIR and narrowband 800 nm (NIR) pulses overlapped at the sample surface with incident angles of 60° and 30°, respectively. The SF signal was then collected in the reflected direction by a spectrograph (Acton SP2300) and a CCD camera (Princeton Instruments PyLoN 1340×400).

**Basic theory of sum-frequency vibrational spectroscopy**

The fundamental theory of SFS is described in Ref. [31]. In the vibrational SFS measurement, a MIR laser beam overlaps spatially and temporally with the up-converting NIR laser beam, and a signal corresponding to the sum of the two input frequencies is detected along the reflected direction. Specifically, when the frequency of the infrared beam approaches a vibrational resonance at the interface, the SF signal can be resonantly enhanced, which is proportional to $\left|\chi_{eff}^{(2)}\right|^2$. Here, $\left|\chi_{eff}^{(2)}\right|^2$ represents the effective nonlinear optical susceptibility, encompassing contributions from both the non-resonant and resonant components of the second-order nonlinearity. According to Ref. [31],

$$S(\omega_{SF} = \omega_{NIR} + \omega_{MIR}) \propto |\chi_{eff}^{(2)}|^2,$$

$$\chi_{eff}^{(2)} = [\hat{e}_{SF} \cdot \overleftrightarrow{L}_{SF}] \cdot \overleftrightarrow{\chi}_{tot}^{(2)} \colon [\hat{e}_{NIR} \cdot \overleftrightarrow{L}_{NIR}][\hat{e}_{MIR} \cdot \overleftrightarrow{L}_{MIR}],$$

$$\chi_{tot}^{(2)} = \overleftrightarrow{\chi}_{NR}^{(2)} + \overleftrightarrow{\chi}_{R}^{(2)},$$

where $\hat{e}_i$ and $\overleftrightarrow{L}_i$ are the unit polarization vector and transmission Fresnel coefficient that acts like a macroscopic local field correction factor at $\omega_i$. When the MIR frequency $\omega_{MIR}$ is near phonon resonances, $\overleftrightarrow{\chi}_R$ has the expression [31]

$$\overleftrightarrow{\chi}_R = \sum_q \frac{\overleftrightarrow{A}_q}{\omega_{MIR} - \omega_q + i\gamma_q},$$

with $\overleftrightarrow{A}_q$, $\omega_q$ and $\gamma_q$ denoting the amplitude, frequency, and damping coefficient of the $q^{\text{th}}$ resonance mode, respectively. For the PPP polarization configuration used in our experiments, where the SF output, NIR input, and MIR input are all $p$-polarized, the effective SF amplitude for an anisotropic surface is described by [31]:

$$A_{PPP} \approx -\cos\beta_{SF}\cos\beta_{NIR}\cos\beta_{MIR}\, L_{xx}(\omega_{SF})L_{xx}(\omega_{NIR})L_{xx}(\omega_{MIR})\cdot A_{xxx}$$
$$-\cos\beta_{SF}\cos\beta_{NIR}\sin\beta_{MIR}\, L_{xx}(\omega_{SF})L_{xx}(\omega_{NIR})L_{zz}(\omega_{MIR})\cdot A_{xxz} +$$
$$\sin\beta_{SF}\sin\beta_{NIR}\sin\beta_{MIR}\, L_{zz}(\omega_{SF})L_{zz}(\omega_{NIR})L_{zz}(\omega_{MIR})\cdot A_{zzz},$$
$$A_{SSP} = cos\,\beta_{MIR}\, L_{yy}(\omega_{SF})L_{yy}(\omega_{NIR})L_{xx}(\omega_{MIR})A_{yyx}$$
$$+\sin\beta_{MIR} L_{yy}(\omega_{SF})L_{yy}(\omega_{NIR})L_{zz}(\omega_{MIR})A_{yyz}.$$

Here, $\beta_{NIR}$, $\beta_{MIR}$ and $\beta_{SF}$ are the NIR input, MIR input and the SF output beam incident angles, respectively. The labels $\{x, y, z\}$ refer to the lab coordinates, with $z$ along the surface normal and $x$ along the beam incident plane.

## Acknowledgements

We thank Peining Li, Xiaosheng Yang, and Weiliang Ma for helpful discussions. J.H.J was supported by the National Key R&D Program of China (2022YFA1404400), the National Natural Science Foundation of China (No. 12125504), the Pioneer Hundred Talents Program of Chinese Academy of Sciences, and the Gusu Leading Innovation Scientists Program of Suzhou City. Q.W, X.W.L, and J.H.J were supported by the start-up funding of the Suzhou Institute for Advanced Research at University of Science and Technology of China. W.T.L was supported by the National Key R&D Program of China (No. 2019YFA0308404), National Natural Science Foundation of China (No. 12250002), and the Science and Technology Commission of Shanghai Municipality (23JC1400400, 22XD1400200, 23DZ2260100). The authors also acknowledge the technical support from the Vacuum Interconnected Nanotech Workstation (Nano-X) at Suzhou Institute of Nano-Tech and Nano-Bionics, Chinese Academy of Sciences.

## Author contributions

J.H.J initiated the project. J.H.J and W.T.L guided the research. Q.W, Z.K.L, L.W.W, Y.L, and J.H.J established the theory. Q.W performed the *ab-initio* calculations. X.L, X.S, and W.T.L performed SFS measurements and analyzed the data. X.W.L carried out the surface treatments and characterization of diamond. C.H.L provided some of the material samples. All the authors contributed to the discussions of the results and the manuscript preparation. J.H.J, Q.W, W.T.L, Z.K.L, X.L, X.W.L, and X.S wrote the manuscript and the supplementary notes.

## Competing Interests

The authors declare that they have no competing financial interests.

## Data availability

All data are available in the manuscript and the Supplementary Information. Additional information is available from the corresponding authors through reasonable request.

## Code availability

We used the commercial software Vienna Ab initio Simulation Package to perform the *ab-initio* calculations in this work. Reasonable request to computation details can be addressed to the corresponding authors.